\documentclass[reprint,aps,prl,superscriptaddress,amsmath,amssymb
]{revtex4-2}

\usepackage{bm}%
\usepackage[colorlinks=true,linkcolor=blue]{hyperref}%
\expandafter\ifx\csname package@font\endcsname\relax\else
 \expandafter\expandafter
 \expandafter\usepackage
 \expandafter\expandafter
 \expandafter{\csname package@font\endcsname}%
\fi
\usepackage{graphicx}
\usepackage{dcolumn}

\RequirePackage[normalem]{ulem}
\RequirePackage{color}

\begin{document}


\title{Coexistent multifractal mesoscopic fluctuations in Integer Quantum Hall Transition and in Orbital Hall Transition}

\author{Nathan L. Pessoa}
\affiliation{Unidade Acad\^emica de Belo Jardim, Universidade Federal Rural de Pernambuco, Belo Jardim-PE 55156-580, Brazil}

\author{Ant\^onio~M.~S.~Mac\^edo}
\affiliation{Laborat\'orio de F\'{\i}sica Te\'orica e Computacional,
Departamento de F\'{\i}sica, Universidade Federal de Pernambuco, Recife-PE 50670-901, Brazil}

\author{Anderson L. R. Barbosa}
\email{anderson.barbosa@ufrpe.br}
\affiliation{Departamento de F\'{\i}sica, Universidade Federal Rural de Pernambuco, Recife - PE, 52171-900, Brazil}

\begin{abstract}
We show that the integer quantum Hall transition in a disordered nanowire with orbital momentum-space texture connected to four terminals is accompanied by an orbital Hall transition. We applied a multifractal detrended fluctuation analysis and found that both conductance fluctuations in the integer quantum Hall transition (IQHT) and orbital-conductance fluctuations in the orbital Hall transition (OHT) display multifractal behavior. We argue that this multifractality is primarily related to disorder, which gives rise to the strong fluctuations that are fingerprints of IQHT and OHT, but is also a consequence of the fact that the nanowire has finite size, which causes a weakening of the multifractality in a certain range of values of disorder strength followed by a new regime of increasing multifractality with increasing disorder strength. {Furthermore, our findings indicate that OHT can bring novel insights to future IQHT analysis.}

\end{abstract}


\maketitle

{\it Introduction} - Mesoscopic fluctuations of quantum transport observables are one the most fundamental features of quantum interference processes in mesoscopic devices at low temperatures \cite{RevModPhys.69.731,RevModPhys.89.015005}.  They were reported in a large variety of systems and phenomena, such as disordered nanowires \cite{PhysRevLett.55.1622,RevModPhys.69.731,PhysRevB.98.155407,PhysRevB.102.115105}, scattering of light fields in disordered media \cite{RevModPhys.89.015005}, integer quantum Hall transition \cite{PhysRevB.49.4679,PhysRevLett.82.4695,PhysRevLett.90.246802}, Majorana wires \cite{PhysRevB.102.195152}, spin Hall effect \cite{PhysRevLett.97.066603,PhysRevLett.98.196601,PhysRevB.93.115120,PhysRevB.102.041107}, orbital Hall effect (OHE) \cite{PhysRevB.108.245105}, thermoelectric effects \cite{KWON2023105691,PhysRevLett.132.246502} and entanglement dynamics \cite{Lim_2024}.

Following unexpected observation of multifractality in the mesoscopic conductance fluctuations (MCF) of single-layer graphene \cite{Amin2018,PhysRevE.104.054129} and HgTe quantum wells \cite{PhysRevLett.131.076301} under low magnetic fields, they reemerged on the scientific scene into the context of complex systems. Multifractality was also reported in mesoscopic spin polarization fluctuations \cite{PhysRevB.107.155432,PhysRevB.110.075421} and mesoscopic thermovoltage fluctuations \cite{PhysRevB.111.L081405}. Multifractality in these systems originates from statistical correlations induced by an externally applied magnetic field in the stochastic process associated with the conductance series \cite{PhysRevE.104.054129}.

When we submit a device to a strong magnetic field, it is driven to the quantum Hall regime, where topologically protected edge states cause quantized Hall conductance plateaus at multiples of $e^2/h$ \cite{PhysRevLett.45.494,PhysRevLett.49.405}. The quantized Hall plateaus result from Landau levels (LLs) that arise in the band structure of the device due to the magnetic field \cite{PhysRevLett.45.494}. The transition between two consecutive quantum Hall plateaus is characterized by strong MCF \cite{PhysRevB.49.4679,PhysRevLett.82.4695,PhysRevLett.90.246802}, and is known as integer quantum Hall transition (IQHT). The latter has many interesting questions about subtle physical phenomena that emerge from the interplay between quantum percolation and localization effects \cite{RevModPhys.67.357,ZIRNBAUER2021168691,DRESSELHAUS2021168676,PhysRevLett.126.076801,PhysRevLett.129.026801,PhysRevB.110.L081112}. {Besides, multifractality was reported in wave functions \cite{PhysRevLett.101.116803} and MCF  \cite{PhysRevLett.128.236803,PhysRevLett.129.186802} in the IQHT.}

Ref. \cite{PhysRevLett.133.146301} reported that OHE accompanies the quantum Hall effect (QHE). More specifically, it was shown that LLs in the quantum Hall regime are orbitally polarized. The OHE converts a longitudinal charge current into a transverse orbital current in materials with momentum-space orbital texture \cite{Go_2021,PhysRevLett.95.066601,PhysRevB.77.165117,PhysRevLett.121.086602,PhysRevResearch.2.013177}. This effect is independent of spin-orbit coupling, opening the range of materials in which it can be observed, in contrast to the spin Hall effect \cite{2023Natur.619...38R,2024NatMa..23.1302L,article}. The OHE has recently been experimentally observed in light metals \cite{2023Natur.619...52C,Hayashi_2023,PhysRevLett.131.156702,PhysRevLett.131.156703,PhysRevApplied.22.064071}, which significantly increases the scientific interest in these materials, as well as in their technological applications \cite{PhysRevResearch.4.033037,PhysRevLett.132.236702,PhysRevLett.128.176601,PhysRevResearch.6.013208,PhysRevLett.134.026303,PhysRevLett.134.036305,PhysRevLett.134.026702,PhysRevLett.134.036304,Seifert_2023,Kumar_2023,articleXu,bdfb41d80ace4bf1b825e5df807ee59f,2023NatPh..19.1855E,PhysRevLett.130.116204,PhysRevResearch.6.013208,PhysRevLett.132.186302,PhysRevB.110.085412,PhysRevLett.133.186302,PhysRevB.110.L140201,veneri2024extrinsicorbitalhalleffect,PhysRevLett.132.246301,cysne2025orbitronicstwodimensionalmaterials,costa2025revealingdominanceorbitalhall,barbosa2025extrinsicorbitalhalleffect,PhysRevB.111.195102,PhysRevB.111.075149,G_bel_2025,https://doi.org/10.1002/aelm.202400554}.

\begin{figure}
    \centering
    \includegraphics[clip, width=0.8\linewidth]{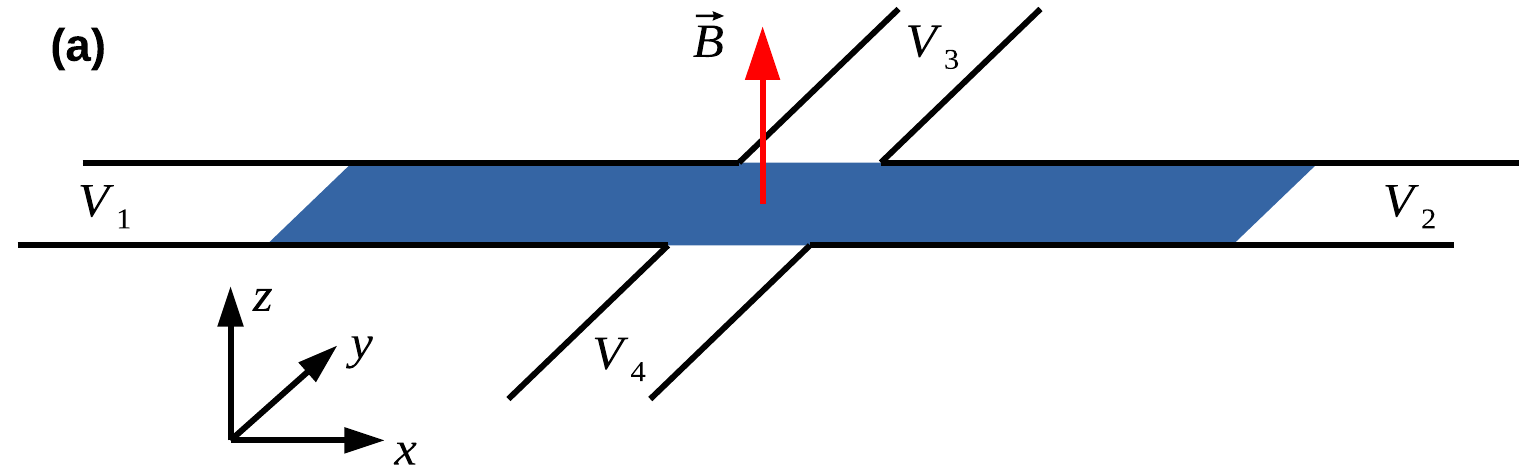}\\
    \includegraphics[clip, width=1.0\linewidth]{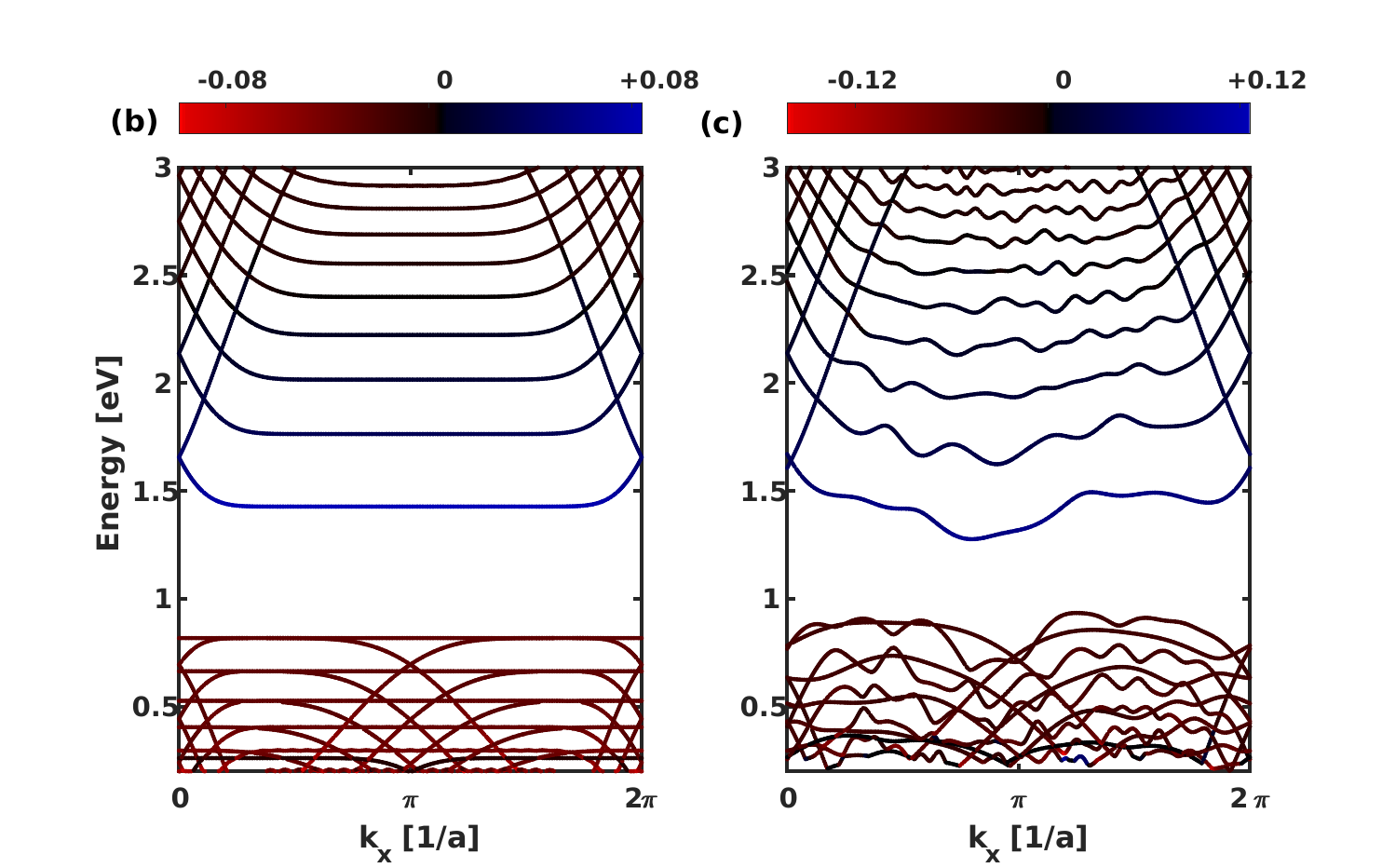}
    \caption{(a) The integer quantum Hall and orbital Hall setup is designed by a nanowire (in blue) submitted to perpendicular magnetical field $\vec{B}$ and connected to four terminals, each submitted to voltage $V_i$ for $i=1,...,4$, where $V_1 >V_2,V_3,V_4$. Down panels are band structures under magnetic flux {$\phi=1/31$} and disorder strength (b) $U=0$ and (c) $U=0.5$ eV. Dispersive edge states are polarized due to the orbital angular momentum $\langle L_z \rangle$ $[\hbar/2]$ (blue, positive; red, negative).}
    \label{HallBar}
\end{figure}

Motivated by \cite{PhysRevLett.128.236803,PhysRevLett.129.186802,PhysRevLett.133.146301}, we study the IQHT in a disordered nanowire with orbital momentum-space texture connected to four terminals under a strong magnetic field, as shown in Fig.~\ref{HallBar}(a). We show that the IQHT is accompanied by an {\it orbital Hall transition} (OHT). More specifically, we show that {\it mesoscopic orbital-conductance fluctuations} in the OHT coexist with the MCF in the IQHT. Furthermore, we interpret both observables as a {\it fictitious time series} where magnetic field and Fermi energy mimic {\it fictitious time}. Using multifractal analysis \cite{KANTELHARDT200287,RAK201848}, we show that {\it mesoscopic orbital-conductance fluctuations} and MCF are similarly multifractal, {which indicates that OHT can bring novel insights to future IQHT analysis} \cite{RevModPhys.67.357,ZIRNBAUER2021168691,DRESSELHAUS2021168676,PhysRevLett.126.076801,PhysRevLett.129.026801,PhysRevB.110.L081112}. Finally, we show that disorder has a strong effect on multifractality.

{\it Microscopic model} -
{The integer quantum Hall and orbital Hall setups are numerically modeled by a centrosymmetric two-dimensional square lattice nanowire with momentum-space orbital texture and without electron-electron and spin-orbit interactions, based on \cite{PhysRevLett.121.086602}}. The nearest-neighbor tight-binding Hamiltonian describing a square lattice with four orbitals (i.e., the $s$ and $p$-orbitals) on each site is \cite{PhysRevLett.121.086602,PhysRevResearch.2.013177}
\begin{eqnarray}
 H&=&\sum_{\langle i,j\rangle\alpha\beta}t_{i\alpha,j\beta}e^{i\theta_{ij}}c_{i\alpha}^{\dagger} c_{j\beta}+ \sum_{i\alpha} \left(E_{i\alpha}+\epsilon_{i}\right) c_{i\alpha}^\dagger c_{i\alpha}, \label{TBH}
\end{eqnarray}
where $\{i,j\}$, and $\{\alpha,\beta\}$ are the unit cell and orbital indices, respectively. The first term represents the nearest-neighbor interaction, where $c_{i\alpha}$ ($c_{i\alpha}^\dagger$) is the annihilation (creation) operator, and $t_{i\alpha,j\beta}$ denotes the hopping integral.  
The perpendicular magnetic field $\vec{B}$ is taken into account by introducing the variables $\theta_{ij}=e/\hbar \int_i^j \vec{A}\cdot d\vec{l}$, where $\vec{A} = \left(-B y,0,0\right)$ and {$\phi=p/q=B a^2/(h/e)$ are the vector potential and the dimensionless magnetic flux, respectively. 
Here, $a$ is the lattice constant and $p$ and $q$ are coprime integers \cite{PhysRevB.14.2239}.} 
The second term contains the on-site energy $E_{i\alpha}$ and the Anderson disorder term $\epsilon_{i}$, which is realized by an electrostatic potential that varies randomly from site to site according to a uniform distribution in the interval $\left(-U/2,U/2\right)$, where $U$ is the disorder strength.
We take the typical Hamiltonian parameters (in eV) $E_s=3.2$, $E_{p_x}=E_{p_y}=E_{p_z}=-0.5$ for on-site energies, $t_s=0.5$, $t_{p\sigma}=0.5$, $t_{p\pi}=0.2$, $t_{sp}=0.5$ for nearest-neighbor hopping integrals \cite{PhysRevLett.121.086602}. 
In our calculations, we used the KWANT software \cite{Groth_2014}.

We applied the above mentioned model to study charge and orbital angular momentum transports through a dirty nanowire with length $L=301a$ and width $W =31a$. 
Fig.~\ref{HallBar}(b) shows the clean (i.e., with disorder strength $U=0$) nanowire's band structure under a magnetic flux {$\phi=p/q=1/31$, which is consistent with $q=31$ LLs in the $s$ orbital \cite{SM}. For $a\approx1$ nm, we have a magnetic field of $B\approx 133$ T, which is very large but of the same order of the magnetic field considered in \cite{PhysRevLett.133.146301}. For an experimental sample where $q$ is large, the magnetic field becomes 5 - 16 T \cite{PhysRevLett.82.4695,PhysRevLett.90.246802,PhysRevLett.129.186802}}. The flat bands corresponding to LLs cluster together in groups and collectively form the bulk energy bands. Conversely, Fig.~\ref{HallBar}(c) shows the dirty (with $U=0.5$ eV) nanowire's band structure with {$\phi=1/31$}. Disorder disrupts the band flatness, causing significant distortion. Additionally, the LLs exhibit orbital polarization, with positive orbital angular momentum $\langle L_z \rangle$ corresponding to $s$-orbital states and negative orbital angular momentum corresponding to $p$-orbital states.
Applying a voltage $V_1 > V_{2,3,4}$, a charge current flows from terminal 1 to terminals 2, 3, and 4 (see Fig.~\ref{HallBar}(a)). Since the LLs are polarized, an orbital current accompanies the charge current. 

From the Landauer-B\"uttiker formalism, the charge and orbital transmission coefficients in the linear regime at low temperatures are calculated from the scattering matrix as
$
T_{ij}^{\eta} =\textbf{Tr}\left[\left(S_{ij}\right)^{\dagger} \mathcal{P}^{L}_{\eta} S_{ij}\right], 
$
where $S_{ij}$ are transmission or reflection matrix blocks of the scattering matrix $S=\left(S_{ij}\right)_{i,j=1,...,4}$ and the matrix $\displaystyle \mathcal{P}^{L}_{\eta}=l^\eta $ is the orbital angular momentum projector \cite{PhysRevB.108.245105}.  The matrices $l^{\eta}$  with $\eta=\lbrace x,y,z\rbrace$ are the orbital angular momentum, and $\eta=0$ stands for the identity matrix in the orbital subspace. Thus, by setting either $\eta=0$ or $\eta=\{x,y,z\}$, the charge and orbital currents can be addressed. 

\begin{figure}
    \centering
    \includegraphics[clip, width=0.9\linewidth]{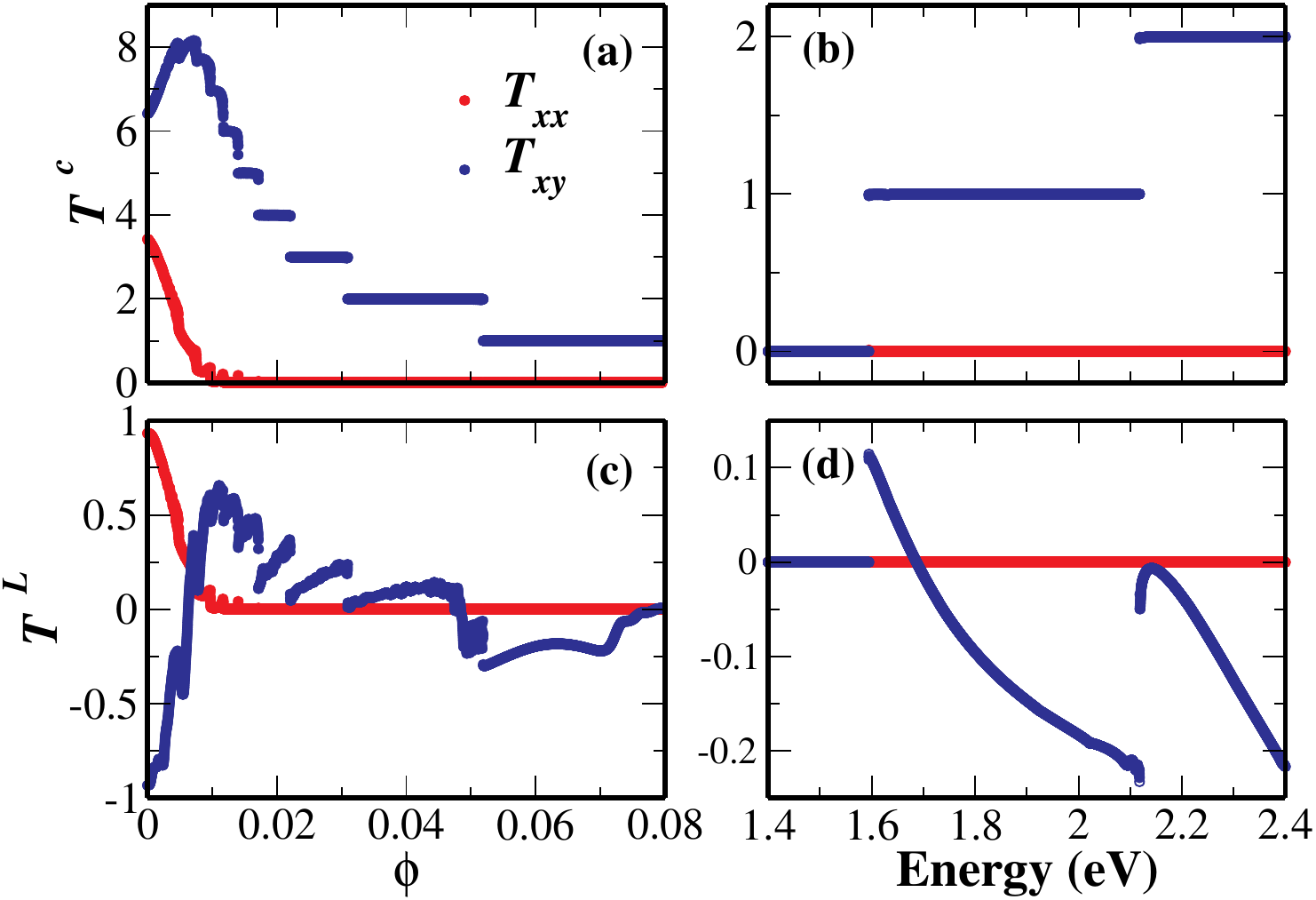}
    \caption{ Longitudinal $T^c_{xx}$ and transversal $T^c_{xy}$ charge transmission coefficients as a function of (a) magnetical flux $\phi$ and (b) Fermi energy through a clean nanowire. Longitudinal $T^L_{xx}$ and transversal $T^L_{xy}$ orbital transmission coefficients as a function of (c) $\phi$ and (d) energy through a clean nanowire. The energy is kept fixed at 2.0~eV in (a) and (c) panels, and $\phi=0.064$ for (b) and (d) panels.}
    \label{ND}
\end{figure}

\begin{figure*}
    \centering
    \includegraphics[clip, width=0.5\linewidth]{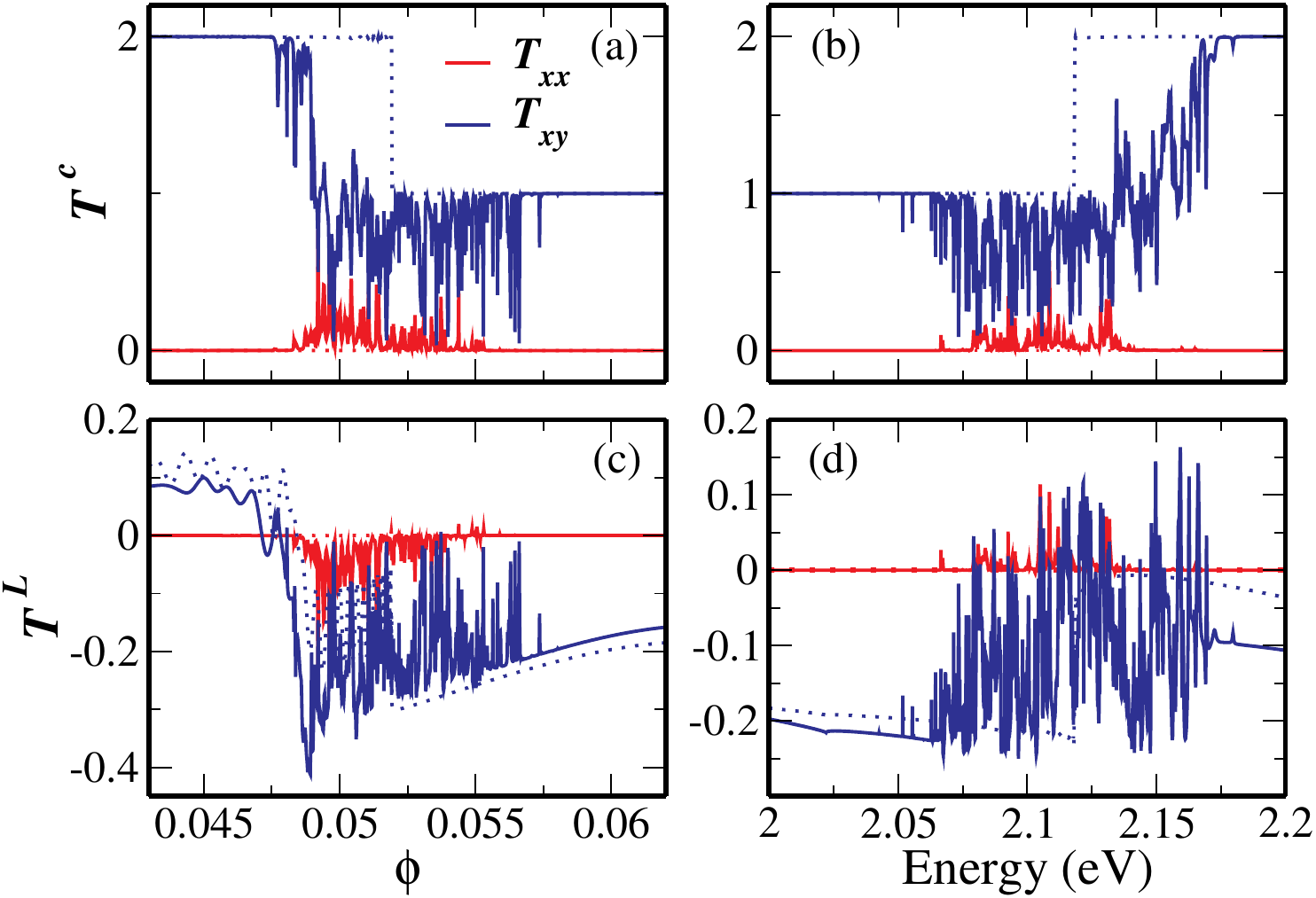}
    \includegraphics[clip, width=0.25\linewidth]{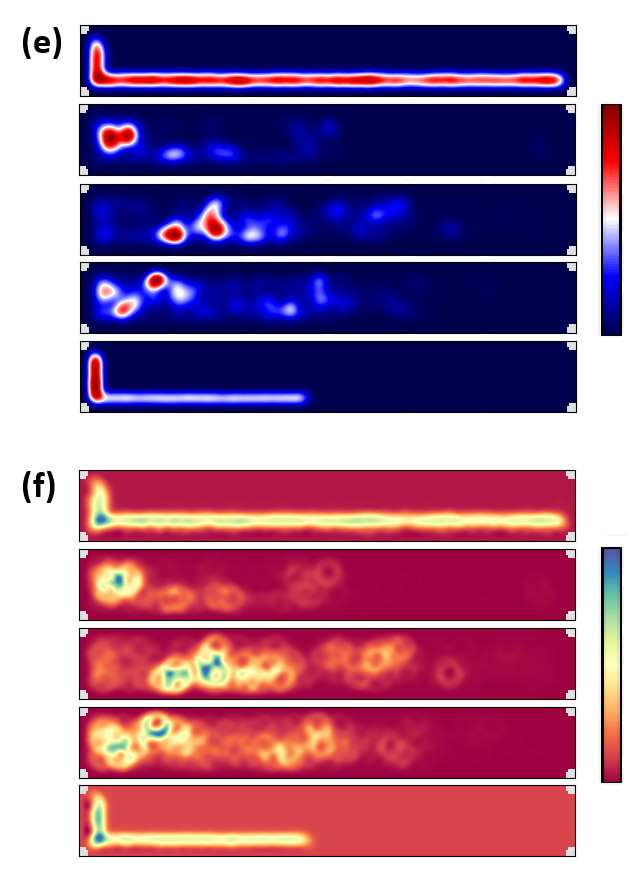}
    \caption{ $T^c_{xx}$ and $T^c_{xy}$ (continuous lines) as a function of (a) magnetical flux $\phi$ and (b) Fermi energy through a dirty nanowire with $U=0.5$~eV. $T^L_{xx}$ and $T^L_{xy}$ (continuous lines) as a function of (c) $\phi$ and (d) energy. The dotted lines are the transmission coefficients from a clean nanowire, Fig.~\ref{ND}. (e) LDOS and (f) ODOS $\langle L_z \rangle$ in the transition between the second and first LL for {$\phi = 0.040$, 0.049, 0.051, 0.053, and 0.064 (from top to bottom)}. The energy is kept fixed at 2.0~eV for (a), (c), (e) and (f) panels and $\phi=0.064$ for (b) and (d) panels.}
    \label{CD}
\end{figure*}

{\it Clean nanowire} - First, we analyze the charge and orbital transmission coefficients through a clean nanowire ($U=0$). Fig.~\ref{ND}(a) shows the longitudinal $T^{c}_{xx} = T_{21}^0$ and transversal $T^{c}_{xy} = T_{31}^0 + T_{41}^0$ charge transmission coefficients as functions of the magnetic flux $\phi$ keeping fixed the Fermi energy at 2.0~eV. $T^c_{xx}$ decreases monotonically to zero by increasing $\phi$. Conversely, $T^{c}_{xy}$ does not have monotonic behavior as a function of $\phi$. After attaining a maximum value, it decreases in quantized Hall plateaus due to the LLs of Fig.~\ref{HallBar}(b). Fig.~\ref{ND}(b) shows $T^{c}_{xx}$ and $T^{c}_{xy}$ as functions of Fermi energy with $\phi=0.064$, where we see that $T^{c}_{xy}$ increases in quantized Hall plateaus, while $T^{c}_{xy}$ remains null when energy increases. 

Fig.~\ref{ND}(c) shows longitudinal $T^{L}_{xx} = T_{21}^z$ and transverse $T^{L}_{xy} = T_{31}^z + T_{41}^z$ orbital transmission coefficients as functions of $\phi$ at energy 2.0~eV, and Fig.~\ref{ND}(d) shows them as functions of Fermi energy at $\phi=0.064$. $T^L_{xx}$ decreases monotonically to zero by increasing $\phi$, similar to the behavior of $T^c_{xx}$. Conversely, $T^{L}_{xy}$ does not display monotonic behavior and changes sign. Besides, $T^{L}_{xy}$ does not exhibit quantized Hall plateaus as a function of either $\phi$ or energy, in contrast to $T^{c}_{xy}$. However, $T^{L}_{xy}$ accompanies $T^{c}_{xy}$ in crossing each LL of Fig.~\ref{HallBar}(b). This confirms that because the LLs are polarized, a transverse orbital transmission accompanies the transverse charge transmission, i.e., OHE accompanies the IQHE, in agreement with \cite{PhysRevLett.133.146301}. The lack of quantized Hall plateaus of orbital transmission coefficients results from the nonconservation of angular momentum in the transport. In contrast, the charge transmission coefficients satisfy the conservation relation $T_{11}^0+T^{c}_{xx}+T^{c}_{xy}=M$, where $M$ is the number of propagating wave modes in the terminals \cite{SM}.

{\it Dirty nanowire} - To study the IQHT, we analyze the transition between the second and first quantized Hall plateaus \cite{PhysRevLett.128.236803,PhysRevLett.129.186802}. Fig.~\ref{CD}(a) shows $T^{c}_{xx}$ and $T^{c}_{xy}$ through a dirty nanowire (continuous lines) as functions of $\phi$ with Fermi energy kept fixed at 2~eV and disorder strength $U = 0.5$~eV. The dotted lines represent the transmission coefficients of the clean nanowire. One sees that $T^c_{xx}$ and $T^c_{xy}$ fluctuate seemingly randomly between plateaus, as reported previously \cite{PhysRevB.49.4679,PhysRevLett.82.4695,PhysRevLett.90.246802}. Furthermore, $T^c_{xx}$ and $T^c_{xy}$ also show similar fluctuations for the first and second Hall plateaus as functions of Fermi energy with $\phi=0.4$ and $U=0.5$~eV, as seen in Fig.~\ref{CD}(b).

Also, $T^{L}_{xx}$ and $T^{L}_{xy}$ through a dirty nanowire are plotted in Fig.~\ref{CD}(c) as continuous lines for the region between the second and first plateaus as functions of $\phi$ with Fermi energy 2~eV and $U=0.5$~eV. Furthermore, in Fig.~\ref{CD}(d), we plotted these coefficients for the region between the first and second plateaus as functions of Fermi energy with $\phi=0.064$ and $U=0.5$~eV. In both cases, $T^{L}_{xx}$ and $T^{L}_{xy}$ fluctuate randomly between the plateaus, indicating the OHT regime. Besides, the mesoscopic fluctuations in the OHT accompany the mesoscopic fluctuations in the IQHT through a dirty nanowire, which is our first primary outcome. By comparing Figs.~\ref{CD}(a,b) to Figs.~\ref{CD}(c,d), 
one sees that, in contrast to the behavior of $T^{c}_{xy}$, $T^{L}_{xy}$ of the dirty nanowire is shifted in relation to $T^{L}_{xy}$ of the clean nanowire, which is one more remarkable difference between $T^{c}_{xy}$ and $T^{L}_{xy}$ \cite{SM}.

To give a better insight into the origins of charge and orbital transmission mesoscopic fluctuations, we plotted the color-coded local density of states (LDOS) in Fig.~\ref{CD}(e) and orbital-polarized density of states $\langle L_z \rangle$ (ODOS) in Fig.~\ref{CD}(f) for {$\phi = 0.040$, 0.049, 0.051, 0.053, and 0.064 (from top to bottom)}. In the top panels of Figs.~\ref{CD}(e,f), we have two LLs; at the bottom, there is only one LL. In both cases, LDOS and ODOS are localized near the lower edge of the dirty nanowire because the only extended states that connect the contacts are edge states. Changing the direction of the magnetic field $B \to -B$, the LDOS and ODOS will be localized near the top edge and $\langle L_z \rangle \to -\langle L_z \rangle$. As the system is driven from a state with two LLs to a state with one LL by increasing $\phi$, the LDOS and ODOS penetrate into the bulk, and a complex spatial pattern develops (middle panels of Figs.~\ref{CD}(e,f)). Thus, we see coherent structures of different sizes forming inside the device. These coherent structures affect the charge and orbital transmissions, giving rise to the mesoscopic fluctuations reported 
in Figs.~\ref{CD}(a-d).

From Figs.~\ref{CD}(e,f), we observe a large range of length scales in the patterns, which is a characteristic of dynamical systems with multiple length (and time) scales. Systems with dynamics across multiple length scales often exhibit multifractal characteristics. To analyze this behavior, we treat both magnetic flux and Fermi energy as independent variables analogous to time parameters. This approach allows us to interpret the longitudinal and transverse charge and orbital transmission coefficients as pseudo-time series, which will be analyzed below in the framework of complex systems. Similar behaviors of LDOS and ODOS are found when Fermi energy increases as the system crosses from states with one LL to ones with two LLs. 

\begin{figure}
    \centering
    \includegraphics[width=\linewidth]{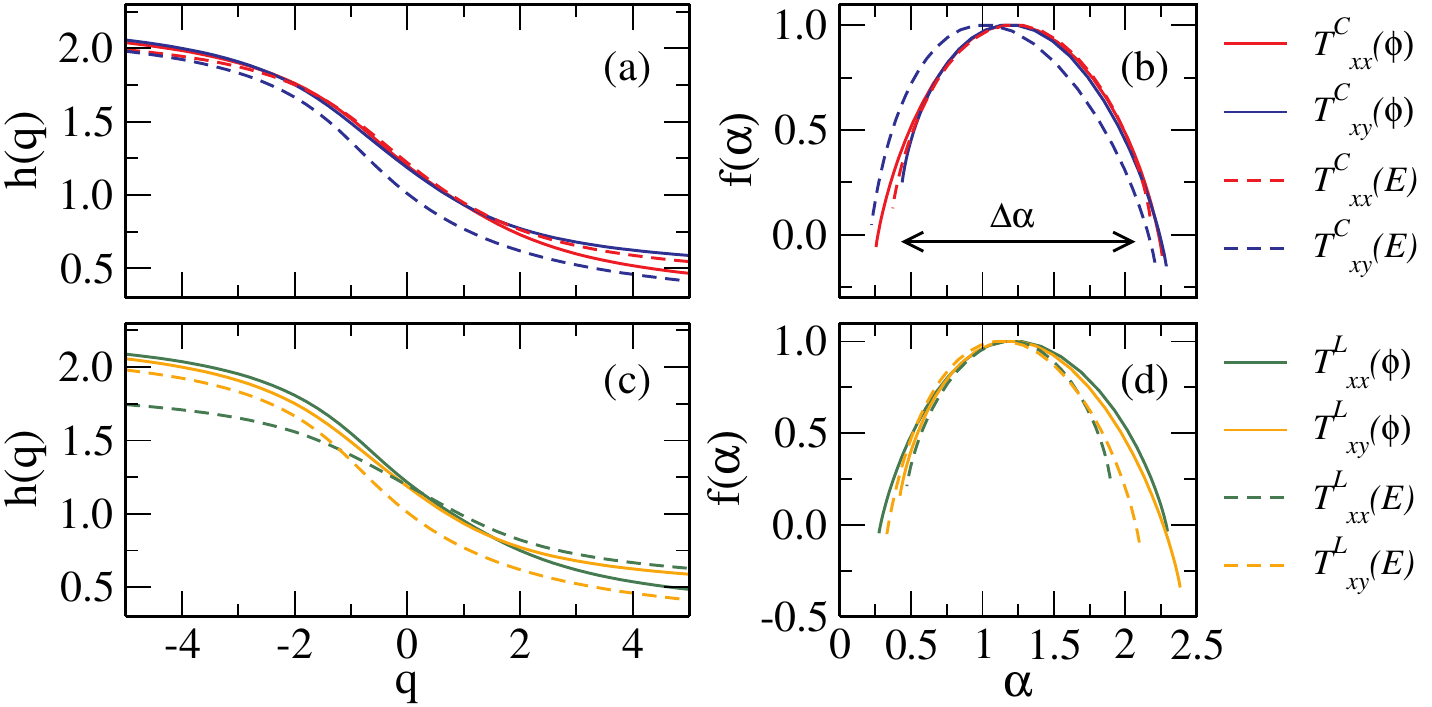}
    \caption{Panels (a,c) show the average Hurst exponent, $h(q)$, and (b,d) the average multifractal singularity spectra, $f(\alpha)$, computed from four independent fictitious time series. Panels (a,b) refer to $T^{c}_{xx,xy}$, while panels (c,d) refer to $T^{L}_{xx,xy}$ as a function of $\phi$ and Fermi energy (legend on the right).The {\it fictitious times series} have 5000 steps range between {$\phi = 0.048,...,0.056$ kept the energy fixed in 2.0 eV, and between energy $0.210,...,0.215$ eV at $\phi = 0.064$}, see Fig.~\ref{CD}(a-d).}
    \label{fig:enter-label}
\end{figure}

{\it Multifractal analysis} - 
To analyze the multifractal behavior of fictitious time series $\{T^{c}_{xx,xy}(t)\}$ and $\{T^{L}_{xx,xy}(t)\}$, where $t$ is the magnetic flux $\phi$ or Fermi energy, we employed the well-known Multifractal Detrended Fluctuation Analysis (MF-DFA) \cite{KANTELHARDT200287,RAK201848}.
In this study, $t=1,...,N$ for all series, and $N=5000$. The series are walk-like, then the first step is to divide them into $N_s=N/s$  non-overlapping  intervals of equal size $s$, and for each $\nu$-th interval, where $\nu=1,..., n_s$,  we perform a linear fit, $T^{c,L}_\nu(t)=a_\nu +b_\nu t$, to the data $T^{c,L}(t)$ and compute the variance:
$
F^2(s,\nu)=\frac{1}{s}\sum_{t=1}^{s} \left[T^{c,L}((\nu-1)s+ t)-T^{c,L}_\nu(t)\right]^{2}.
$
Finally, we compute the $q$-th order fluctuation function defined by 
$
F_q(s)= \langle [F^2(s,\nu)]^{q/2}\rangle^{1/q}=\left[\frac{1}{N_s}\sum_{\nu=1}^{N_s}\left[F^2(s,\nu)\right]^{q/2}\right]^{1/q}.
$
The multifractal scaling exponent or generalized Hurst exponent,  $h(q)$, is then defined  by the following scaling relationship $
F_q(s)\sim s^{h(q)}$. Our analysis considers $q$ values ranging from $-4$ to $4$ in steps of $0.2$ \cite{SM}. The series is classified as multifractal if  $h(q)$ varies with $q$ and monofractal if  $h(q)$ is independent of $q$. We can also distinguish the behavior of the series between multifractal and monofractal through the singularity spectrum, which is defined by employing a Legendre transformation to the function $\tau(q)=qh(q)-1$, which leads us to $f(\alpha)=\alpha q - \tau(q)$, where $\alpha=d \tau / dq$. The series is multifractal if the singularity spectrum $f(\alpha)$ has a significant width $\Delta\alpha =\alpha_{\rm{max}}-\alpha_{\rm{min}}$, and monofractal if $f(\alpha)$ is narrow, i.e., $\Delta\alpha \rightarrow 0$.

Fig.~\ref{fig:enter-label}(a) shows the average generalized Hurst exponent $h(q)$ as a function of $q$ obtained from four independent fictitious time series $\{T^{c}_{xx,xy}(t)\}$ for dirty nanowires with $U = 1.5$ eV. We see that $h(q)$ varies significantly with $q$, indicating that the $\{T^{c}_{xx,xy}(t)\}$ are multifractal.  Fig.~\ref{fig:enter-label}(b) supports these results, showing that the average multifractal singularity spectra $f(\alpha)$ are considerably wide ($\Delta\alpha \simeq 2$). Notably, $h(q)$ and $f(\alpha)$ calculated from $T^{c}_{xx}(t)$ and $T^{c}_{xy}(t)$ show similar multifractal behavior. These results agree with \cite{PhysRevLett.128.236803,PhysRevLett.129.186802}, which first reported the multifractal mesoscopic fluctuations in the IQHT.

Furthermore, Fig.~\ref{fig:enter-label}(c) shows the average value of $h(q)$ as a function of $q$ obtained from four independent fictitious time series $\{T^{L}_{xx,xy}(t)\}$ for the dirty nanowires with $U = 1.5$ eV. As we can clearly see, they show a large variation with $q$, indicating that $\{T^{L}_{xx,xy}(t)\}$ are multifractal. Besides, this result is supported by Fig.~\ref{fig:enter-label}(b), which shows that the average multifractal singularity spectra $f(\alpha)$ have a large width $\Delta\alpha \simeq 2$. Notably,  $h(q)$ and $f(\alpha)$ calculated for $\{T^{L}_{xx,xy}(t)\}$ show similar multifractal behavior to the ones calculated for $T^{c}_{xx,xy}(t)$. This indicates that OHT cannot be neglected in the IQHT analysis, which is our second primary outcome.

\begin{figure}
    \centering
    \includegraphics[width=0.9\linewidth]{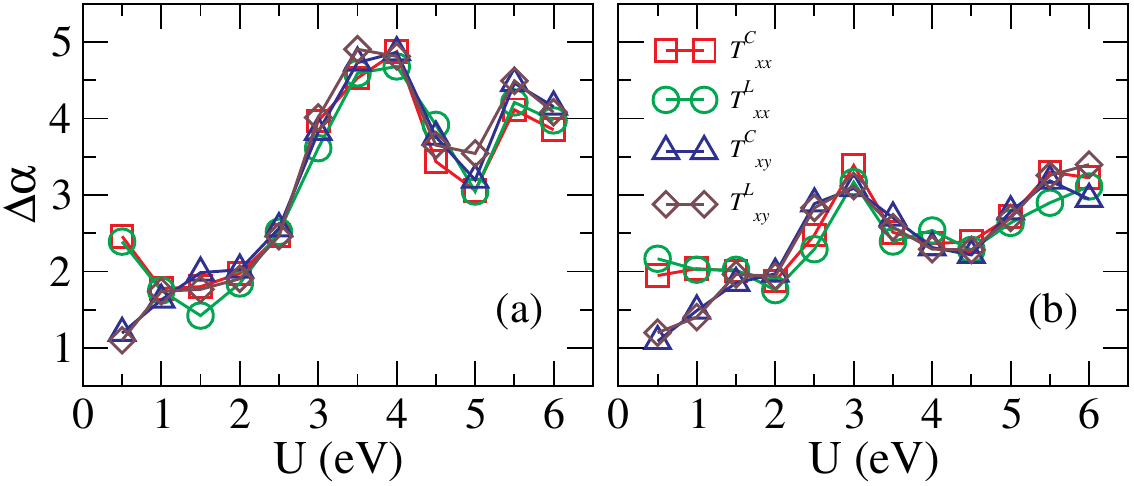}
    \caption{Width of the mean multifractal singularity spectra, $\Delta\alpha$, as a function of the disorder strength $U$ calculated from four independents  fictitious time series (a) $T^{c,L}_{xx,xy}(E)$ as a function of the Fermi energy holding $\phi = 0.064$ fixed and (b) $T^{c,L}_{xx,xy}(\phi)$ as a function of $\phi$ at energy 2.0 eV.}
    \label{Deltaalpha}
\end{figure}

Finally, we analyze the effect of disorder on the multifractal behavior of mesoscopic fluctuations in IQHT and OHT. Fig.~\ref{Deltaalpha} shows the width of the average multifractal singularity spectra $\Delta\alpha$, as a function of the disorder strength $U$. Values on panel (a) are obtained from $T^{c,L}_{xx,xy}$ as functions of Fermi energy with $\phi = 0.064$ and on panel (b) from $T^{c,L}_{xx,xy}(\phi)$ as functions of $\phi$ with Fermi energy 2.0 eV. In panels (a,b), we see that even for weak disorder, namely $U=0.5$ eV, $\Delta\alpha \simeq 1$, a large enough value to indicate multifractity of a time series. Furthermore, $\Delta\alpha$ increases when $U$ increases, attaining its maximum value around $U=3.5$ eV, which indicates a strong dependence of multifractal behavior with the disorder. The multifractal spectrum width $\Delta\alpha$ reaches a maximum value of approximately 5 in panel (a), compared to approximately 3 in panel (b). This difference can be attributted to the fact the dirty nanowires of the panel (a) are submitted to $\phi = 0.064$, while the ones of the panel (b) are submitted to a magnetic flux between $0.048$ and $0.056$, indicating that the multifractal behavior of mesoscopic fluctuations are strongly dependent on the magnetic field intensity.

The multifractal character of charge and orbital conductance fluctuations can be attributed to two main aspects. First, as mentioned before, multifractity is a key feature of the fluctuations observed in systems with multiscale dynamics, as a four-terminal disordered mesoscopic device in the IQHT (and OHT) regime; however, the fluctuations are  sensitive to the size of the specimen, and that is the second main reason for the observed multifractality. This explains the existence of two peaks in Figs.~\ref{Deltaalpha}(a,b), showing that disorder is not the only mechanism that gives rise to the multifractal behavior, but also it is a finite-size effect.

{\it Conclusions} - In this work, we have shown that the IQHT in a disordered nanowire with orbital momentum-space texture is accompanied by an OHT. Our quantum transport calculations revealed that mesoscopic orbital-conductance fluctuations in the OHT coexist with MCF in the IQHT. By treating magnetic field and Fermi energy as fictitious time variables, we conducted a multifractal detrended fluctuation analysis that showed both phenomena exhibit similar multifractal behavior, with significant dependence on disorder strength. The coexistence of these multifractal signals highlights that orbital angular momentum transport is an intrinsic feature of the quantum Hall regime, not merely a secondary effect. Our findings confirm and extend the recent observation that LLs in quantum Hall systems are orbitally polarized \cite{PhysRevLett.133.146301}, generating orbital-polarized edge currents alongside the well-established charge currents. The similar multifractal spectra of charge and orbital conductance fluctuations suggest a common underlying critical mechanism, {indicating that OHT can bring novel insights to IQHT} \cite{RevModPhys.67.357,PhysRevLett.101.116803,ZIRNBAUER2021168691,DRESSELHAUS2021168676,PhysRevLett.126.076801,PhysRevLett.129.026801,PhysRevB.110.L081112}. We also identified a non-monotonic relationship between disorder strength and multifractality, where finite-size effects in the nanowire lead to a suppression of multifractal behavior at intermediate disorder strengths, followed by enhanced multifractality at higher disorder values. This result provides insight into the interplay between localization, system size, and critical fluctuations near the quantum Hall transition.

These findings open several promising research directions. Future theoretical work could develop a unified scaling theory incorporating both charge and orbital degrees of freedom at criticality. Experimentally, the multifractal nature of orbital conductance fluctuations could be probed through techniques that detect orbital magnetization, such as magneto-optical Kerr effect \cite{PhysRevLett.131.156702} or through orbital-to-spin conversion processes \cite{PhysRevResearch.4.033037}. The quadratic magnetic field dependence of orbital Hall resistivity, distinct from the linear scaling of charge Hall resistivity, provides a potential signature to distinguish these effects in high-field measurements. Additionally, our methodology connecting multifractal analysis with quantum transport could be extended to other topological systems, potentially revealing similar complex statistical patterns in quantum spin Hall transitions \cite{PhysRevLett.95.226801,PhysRevLett.115.136804,PhysRevLett.130.196401} or anomalous Hall systems \cite{PhysRevLett.134.026702}. The intersection of quantum topology, disorder-induced criticality, and orbital physics demonstrated here represents a rich framework for exploring fundamental aspects of quantum phase transitions and their potential applications in orbitronics.

\begin{acknowledgments}
This work was supported in part by the following Brazilian agencies: Conselho Nacional de Desenvolvimento Cient\'ifico e Tecnol\'ogico (CNPq), under Grant Numbers 309457/2021 (ALRB) and 307626/2022-9 (AMSM), Coordena\c c\~ao de Aperfei\c coamento de Pessoal de N\'ivel Superior (CAPES), Fundação de Amparo à Ciência e Tecnologia de Pernambuco (FACEPE), and INCT of Spintronics and Advanced Magnetic Nanostructures (INCT-SpinNanoMag), Grant No. CNPq 406836/2022-1.
\end{acknowledgments}

\bibliography{references}

\end{document}